\documentclass[conference]{IEEEtran}
\IEEEoverridecommandlockouts

\usepackage{float}
\usepackage{hyperref}

\usepackage{cite}
\usepackage{amsmath,amssymb,amsfonts}
\usepackage{algorithm}
\usepackage{algorithmicx}
\usepackage{graphicx}
\usepackage{textcomp}
\usepackage[OT1]{fontenc}
\usepackage{xcolor}

\usepackage{threeparttable}
\usepackage{tikz}

\usepackage{subfigure}

\usepackage[
	n,
	operators,
	advantage,
	sets,
	adversary,
	landau,
	probability,
	notions,	
	logic,
	ff,
	mm,
	primitives,
	events,
	complexity,
	asymptotics,
	keys]{cryptocode}

\def\BibTeX{{\rm B\kern-.05em{\sc i\kern-.025em b}\kern-.08em
    T\kern-.1667em\lower.7ex\hbox{E}\kern-.125emX}}
\begin{document}

\title{IBRS: An Efficient Identity-based Batch Verification Scheme for VANETs Based on Ring Signature}

\author{\IEEEauthorblockN{Feng Liu, Qi Wang}
\IEEEauthorblockA{\textit{Department of Computer Science and Engineering} \\
\textit{Southern University of Science and Technology}\\
Shenzhen, Guangdong 518055, China \\
liuf2017@mail.sustech.edu.cn, wangqi@sustech.edu.cn}
}
\maketitle

\begin{abstract}

Vehicular ad-hoc networks (VANETs) are one of the most important components in an Intelligent Transportation System (ITS), which aims to provide information communication between vehicles.
A safety-critical vehicular communication requires security, privacy, auditability and efficiency.
To satisfy these requirements simultaneously, several conditional privacy-preserving authentication schemes are proposed by employing ring signature.
However, these methods have been paid too little attention to the issues like \textit{how to choose the valid ring members} or \textit{how to set up a ring}.
In this paper, we introduce an efficient conditional privacy-preserving scheme which provides an appropriate approach establishing the list of ring members.
Moreover, our proposed scheme also supports batch verification to significantly reduce the computational cost.
According to the analysis of security, our scheme is sufficiently resistant against several common attacks in VANETs.
The performance results show that the proposed scheme is efficient and practical with both low computation and communication cost.

\end{abstract}

\begin{IEEEkeywords}
VANETs, ITS, ring signature, conditional privacy, batch verification
\end{IEEEkeywords}

\section{Introduction}\label{I}
Vehicular Ad-hoc Networks (VANETs) as a special case of Mobile Ad-hoc Networks (MANETs) optimized for vehicular environments play an important role in Intelligent Transportation Systems (ITS).
In a typical scenario of ITS, each vehicle broadcast traffic-related information, such as its speed, position, the road condition, etc. via VANETs.
After receiving these broadcast messages, vehicles analyze and extract meaningful information to drivers, or take corresponding control actions in some emergencies.
In this way, road safety and efficiency will be greatly enhanced, and this is essential for automated vehicles.
However, due to the high demand for road safety features, to design a practical protocol for VANETs is highly nontrivial.
Most of the proposed schemes are built based on the IEEE 802.11p standard.

\begin{figure}[htbp]
    \centering
    \includegraphics[scale=0.45]{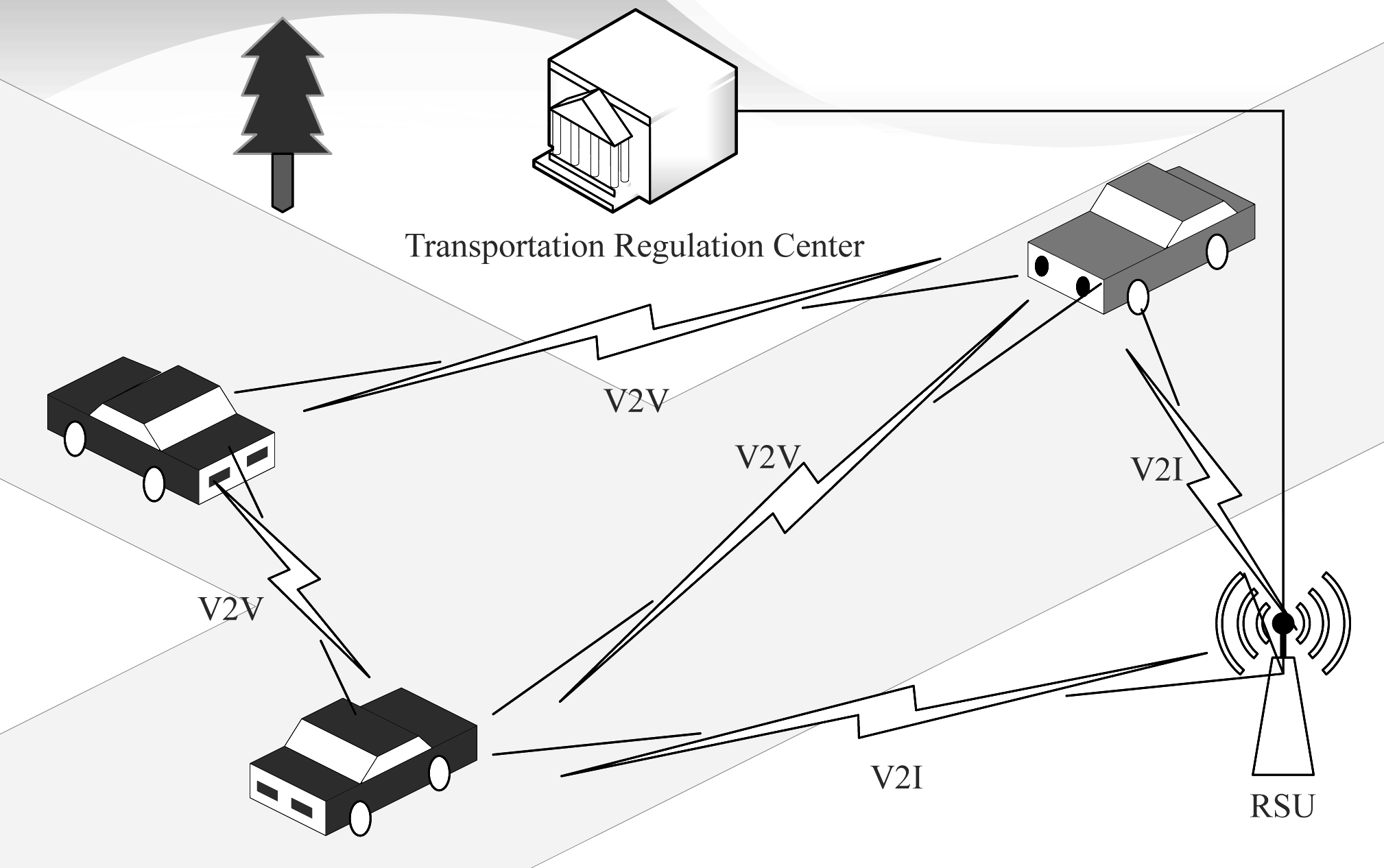}
    \caption{VANETs overview}
    \label{fig:01}
\end{figure}

In IEEE 802.11p, the participants in the road are classified into two categories, i.e., On-Board Units (OBUs) and Road-Side Units (RSUs).
Each vehicle is equipped with an OBU for broadcasting messages and handling the received messages.
The RSUs are usually fixed along roads as the base stations to provide Internet access and extra road information for vehicles.
Therefore, VANETs provide two different types of communication, namely, vehicle-to-infrastructure (V2I) and vehicle-to-vehicle (V2V) as shown in Fig.~\ref{fig:01}.
Because the broadcast messages are necessary in important applications, like collision avoidance, traffic optimization, we suppose that these messages have no necessary relevance to vehicles' identities.
As a result, drivers can be reminded by receiving the broadcast messages from other vehicles or RSUs in VANETs.
In practice, a trusted party, the Transportation Regulation Center (TRC), is needed to administrate the whole network.
RSUs can connect with TRC for obtaining extra information.

However, because of the open environment of VANETs, an attacker could send a forged message to confuse other drivers, which may further cause potential traffic hazards.
To achieve road safety, it is essential to authenticate the validity of a message.
The first serious discussions of road safety emerged in 2002 and utilized digital signatures~\cite{el2002security}.
After that, a considerable amount of work has been done based on different digital signature schemes~\cite{petit2014pseudonym}.

When applying digital signature schemes in VANETs, upon each message will attach some extra information including signature, signer's certificate and so on.
This extra information may be used to link to the driver's true identity, and this may lead to privacy disclosure.
Therefore, \textit{how to keep the anonymity of senders} while the message can be authenticated by verifiers becomes another essential issue in VANETs.
One common approach is to replace the true identity with a random-like string called pseudonyms.
On the other hand, in some specific scenarios such as a traffic accident, there should be a possibility to reveal the true identity of senders according to the attached information.
To this end, a function of auditability should be provided.
Thus, authentication, privacy, and auditability are three basic requirements when designing a feasible scheme for communication in VANETs.
Besides, due to the limited computation and storage capability of both OBUs and RSUs, efficiency should also be considered in VANETs.

In this paper, we propose a hybrid scheme, which employs ring signatures in VANETs with batch verification mode, and hereby efficiently enables auditability for ring members. 
More precisely, our contributions are summarized in the following:

\begin{itemize} 
 \item We propose a novel scheme for VANETs based on identity-based ring signature, where the procedure of creating a ring is restricted to make ring members auditable.

 \item We provide a batch mode for message verification. As indicated by performance, this makes our scheme highly efficient. To the best of our knowledge, this is the first attempt that applies ring batch verification in VANETs.

 \item We give security analysis of the proposed scheme and implement our scheme in the Raspberry Pi 3b+ platform. 
\end{itemize}

The remainder of this paper is organized as follows.
In Section~\ref{II}, some related work is presented to balance authentication and privacy in VANETs.
Section~\ref{III} introduces the system model and some preliminary cryptographic primitives.
In Section~\ref{IV}, a description of our schemes is given in detail.
After that, security analysis and performance analysis are provided in Section~\ref{V} and Section~\ref{VI} respectively.
Finally, Section~\ref{VII} concludes this paper and proposes some potential future work.

\section{Related work}\label{II}
Numerous studies have attempted to employ pseudonym schemes to assure authentication and privacy simultaneously.
specifically, the cryptography tools utilized include public key infrastructure (PKI), identity-based cryptography (IBC), group signature, ring signature and so on (for recent surveys, see~\cite{petit2014pseudonym,ali2019authentication}).

At the early stages of the study, PKI is most widely used in VANETs.
In these schemes based on PKI, as a trusted party, Certificate Authority (CA) is needed.
Each vehicle broadcasts messages attached to the corresponding signatures and public-key certificates.
Taking the SeVeCom project~\cite{wiedersheim2009sevecom} as an example, the elliptic curve digital signature algorithm (ECDSA) is utilized to assure efficiency.
A pseudonym consists of two parts: a short-term key and its corresponding certificate.
Since the use of certificates increases the communication overhead, it was alternatively suggested to employ IBC instead of certificates~\cite{el2002security,kamat2006identity}.

Similarly, IBC-based schemes also adopt a set of short-term public keys to form vehicles' pseudonyms, while the procedure of pseudonym issuance differs.
Note that in IBC-based schemes, a new entity---private key generator (PKG)--- is introduced without the existence of CA.
Thus, certificates are not attached when broadcasting messages in these schemes, and the communication overhead is thereby decreased.

There exists a major problem in both PKI-based and IBC-based pseudonym schemes: pseudonym change since it is not sufficient to use a single pseudonym to preserve vehicles' privacy.
In these two kinds of schemes, using a single pseudonym is not sufficient to preserve the vehicle's privacy.
As a simple setting, each vehicle is equipped with a set of public keys, each of which can be viewed as an unlinkable pseudonym, and expires after a fixed amount of time.
Wiedersheim et al.~\cite{wiedersheim2010privacy} pointed out that simple pseudonym change is not enough to preserve privacy.
There have been attempts on the strategy of pseudonym change, e.g., mix-zone-based~\cite{ying2013dynamic} and mix-context-based~\cite{gerlach2007privacy}.
However, it is still mysterious to formalize the relationship between pseudonym change strategies and privacy level~\cite{petit2014pseudonym,boualouache2017survey}.

The issue of pseudonym change can be eliminated in those schemes based on group signature and ring signature~\cite{petit2014pseudonym,calandriello2007efficient}, since messages are signed under the identity of a certain group rather than a single vehicle's pseudonym. 
Group signature-based schemes allow a vehicle to sign a message anonymously on behalf of the group.
In group signatures, a special entity called group manager can reveal any signer's real identity from the corresponding signature.
Till now, there has been little agreement on the choice of group manager~\cite{petit2014pseudonym}.
As the administrator of the group, the group manager has the privilege to add or delete a group member.
It is then straightforward to achieve auditability of group members by the group manager.
It was suggested that RSUs serve as group managers~\cite{park2011rsu,zhang2009scalable}.
However, RSUs are vulnerable to some extent, and this setting is not sufficient to guarantee group members' privacy

\begin{figure}[htbp]
   \centering
   \includegraphics[scale=0.2]{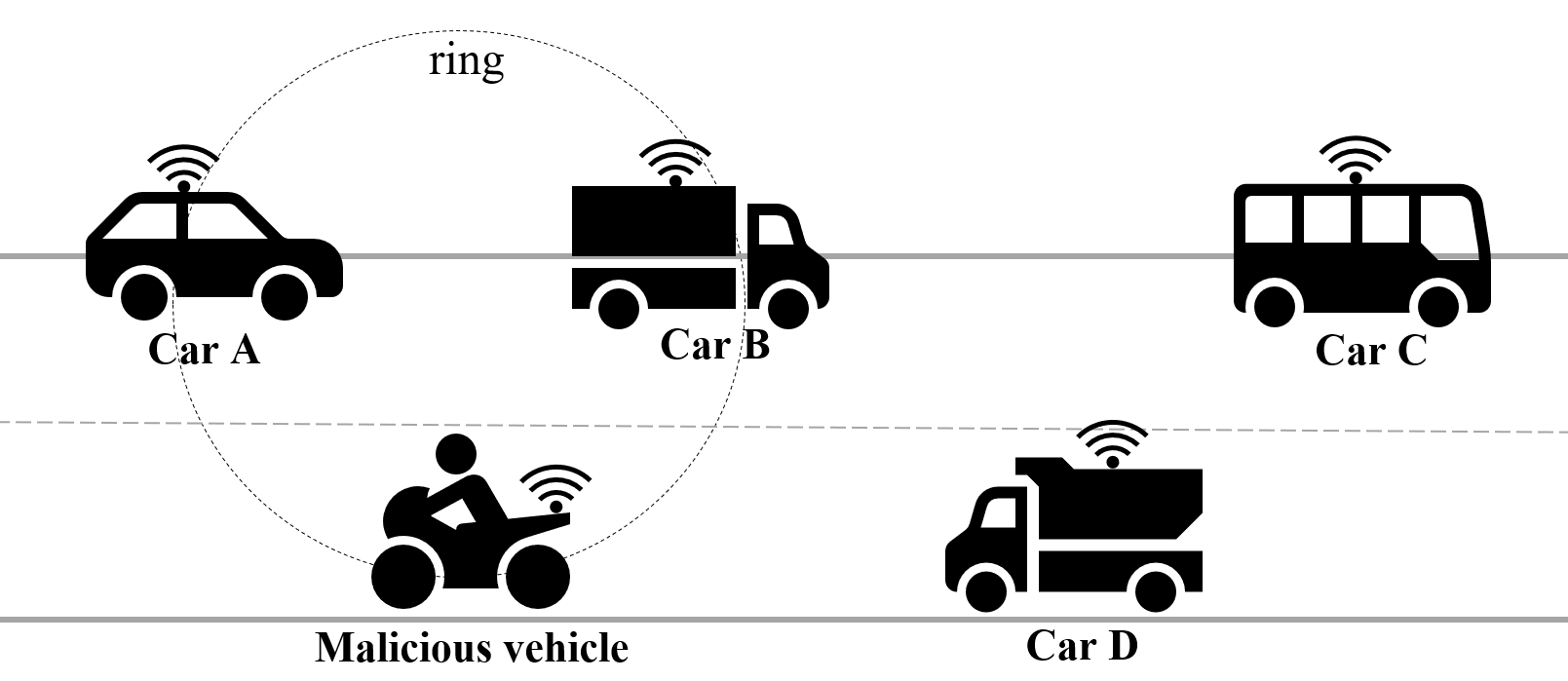}
   \caption{A negative case in ring signature-based schemes}
   \label{fig:02}
\end{figure}

In comparison, ring signature-based schemes~\cite{zeng2015privacy,jiang2014anonymous,zeng2018concurrently} further removes group managers by involving a set of different vehicles' public keys as the ``group'' (ring).
In existing ring signature-based schemes, e.g.,~\cite{zeng2015privacy}, each vehicle can collect other vehicles' public keys on roads and thereby checks the validity of ring members before verification.
Unfortunately, this would lead to verification failure when a malicious vehicle broadcasts an invalid public key.
As depicted in Fig.~\ref{fig:02}, due to the existence of a malicious vehicle in the ring, the generated signature by the whole ring would be rejected by other vehicles.
In existing such schemes, privacy is the main focus while the lack of auditiability of ring members is still a problem~\cite{chaurasia2011conditional,zeng2015privacy}.

Notably, Petit et al.~\cite{petit2014pseudonym} emphasize that these categories are not hard-edged so that several recent works combine different techniques from the previous categories.
Survey~\cite{ali2019authentication} listed these hybrid schemes and discussed their security and efficiency in performance.
These results pointed out that recently proposed schemes attempt to apply the batch verification of signatures into the verification procedure, which can greatly reduce the computation cost comparing with single verification.

\section{Preliminaries}\label{III}

In this section, we briefly introduce the system model, and several cryptographic primitives, including bilinear pairings, identity-based encryption, and identity-based ring signature.

\subsection{System model}

Generally, our VANETs model consists of four main entities: the Transportation Regulation Center (TRC), Law Enforcement Authorities (LEA), RSUs and vehicles equipped with OBUs.
The detailed description of these entities is listed as follows.

\begin{itemize}
    \item \textbf{TRC:} The TRC is a fully trusted party in the VANETs system with sufficient computation and storage capabilities.
    TRC takes charge of system initialization and registration of other entities like RSUs and vehicles.
    We assume that TRC can establish a secure channel with each RSU.
    Besides, TRC is also responsible for identity reveal and OBU revocation.
    Typically, to record the identities of those invalid vehicles, TRC manages a revocation list.
    \item \textbf{LEA:} LEA is the agency for ensuring the accountability of vehicles.
    In other words, when a vehicle broadcasts fake messages anonymously on purpose, the LEA can reveal the identity of the vehicle with the help of TRC.
    \item \textbf{RSUs:} An RSU usually plays an auxiliary role between TRC and vehicles.
    Namely, RSUs can communicate with TRC through wired or wireless networks and broadcast messages to vehicles in a restricted region.
    In our system, RSUs can obtain a revocation list from TRC periodically and delivery ring-member lists to vehicles. 
    \item \textbf{Vehicles:} Each vehicle in this system is equipped with a communication device called OBU.
    We assume that each OBU contains hardware security modules (HSM), and the required cryptographic operations are executed inside HSMs.
\end{itemize}

\subsection{Bilinear pairings}\label{III-B}

Bilinear pairings have been widely used to design various cryptographic schemes over the last two decades~\cite{boneh2001identity,chow2005efficient,zhang2002id}.
Let $\mathbb{G}_1$, $\mathbb{G}_2$, $\mathbb{G}_T$ be three cyclic groups of the same prime order $q$.
Assume that the discrete logarithm problem in $\mathbb{G}_1$, $\mathbb{G}_2$ and $\mathbb{G}_T$ is hard.
Let $\hat{e}:\mathbb{G}_1 \times \mathbb{G}_2\rightarrow \mathbb{G}_T$ be a bilinear pairing with the following properties:

\begin{itemize}
  \item[1.] Bilinearity: $\forall P\in \mathbb{G}_1$, $\forall Q\in\mathbb{G}_2$ and $\forall a,b\in \mathbb{Z}_q^*$, $e(aP,bQ)=e(P,Q)^{ab}$;
  \item[2.] Non-degenerate: $\exists P\in \mathbb{G}_1$, $\exists Q\in \mathbb{G}_2$ such that $e(P,Q)\neq1$;
  \item[3.] Computability: $\forall g_1\in \mathbb{G}_1$,$\forall g_2\in \mathbb{G}_2$, there is an efficient algorithm to compute $\hat{e}(g_1,g_2)$.
\end{itemize}

There are two hard-problem assumptions in bilinear pairings, i.e., Computational Bilinear Diffie-Hellman (CBDH) Problem and Decisional Bilinear Diffie-Hellman (DBDH) Problem, as described in the following.
\begin{itemize}
    \item CBDH: Given $P\in\mathbb{G}_1$, $aQ$,$bQ$,$cQ\in(\mathbb{G}_2)^3$, where $a,b,c\in_R(\mathbb{Z}_q^*)^3$, it is difficult to calculate $\hat{e}(P,Q)^{abc}$.
    \item DBDH: Given $P\in\mathbb{G}_1$, $aQ$,$bQ$,$cQ\in(\mathbb{G}_2)^3$, $h\in\mathbb{G}_T$, where $a$,$b$,$c\in_R(\mathbb{Z}_q^*)^3$, it is difficult to determine whether or not $h=\hat{e}(P,Q)^{abc}\ \mathsf{mod}\ q$.
\end{itemize}

\subsection{Identity-based cryptography}\label{III-C}

Bilinear pairings are used to construct identity-based encryption and signature schemes.
Compared to traditional PKI, identity-based cryptography avoids CA since each user's public key can be automatically derived from the corresponding identity(e.g., user's phone number, email address)
In general, a common Identity-based cryptosystem contains two basic algorithms.

\begin{itemize}
    \item [1.] $\mathsf{Setup}(1^\kappa)\rightarrow \mathsf{pp}$: Taken the input of security parameter $\kappa$, the algorithm $\mathsf{Setup}(1^\kappa)$ first chooses a master secret key $s\in_R\mathbb{Z}_q^*$, and then outputs the public parameter $pp=\{\mathbb{G}_1,\mathbb{G}_2,\mathbb{G}_T,P,Q,PK_1,PK_2,q,\hat{e},H_1,H_2\}$, where $\mathbb{G}_1,\mathbb{G}_2,\mathbb{G}_T,q,\hat{e}$ are described in Section~\ref{III-B}.
    $P$ is the generator of $\mathbb{G}_1$, $Q$ is the generator of $\mathbb{G}_2$, and $PK_1=s\cdot P$, $PK_2=s\cdot Q$.
    $H_1$, $H_2$ and $H$ are three cryptographic hash function where $H_1:\{0,1\}^*\rightarrow\mathbb{G}_1$, $H_2:\{0,1\}^*\rightarrow\mathbb{G}_2$, $H:\{0,1\}^*\rightarrow\mathbb{Z}_q^*$.
    \item [2.] $\mathsf{KeyGen}(\mathsf{ID}_i)\rightarrow \{pk_i,sk_i\}$: When user $i$ wants to obtain his/her public key and private key from the system, he/she first sends the specific identity $\mathsf{ID}_i$ to system. After authentication is accepted, the system runs $\mathsf{KeyGen}(\mathsf{ID}_i)$ to derive the user's public key $pk_i$ and private key $sk_i$, where $pk_i=H_1(ID_i)$ and $sk_i=s\cdot pk_i$.
\end{itemize}

The first practical identity-based encryption scheme~\cite{boneh2001identity} was proposed in 2001 by Boneh and Franklin.
In the rest of this paper, we use $\mathsf{Enc}_{pk_i}(\cdot)$ and $\mathsf{Dec}_{sk_i}(\cdot)$ to denote the variant of Identity-based encryption and decryption algorithms in~\cite{boneh2001identity} respectively.

Furthermore, an identity-based ring signature is also used in our scheme, which requires verifiers to verify messages through a specific set of signers.
Being different from the traditional signature, an identity-based ring signature scheme requires verifier to verify messages through a specific set of signers.

We adopt the identity-based ring signature in~\cite{chow2005efficient} in our proposed scheme.
There are two essential algorithms, i.e., signature algorithm $\mathsf{Sign}_{sk_i}(m,L)\rightarrow \sigma$ and verification algorithm $\mathsf{Verify}_{L}(\sigma,m)\rightarrow 0/1$, where $m$ and $L$ denote the message and the ring-member list, respectively.

\section{The proposed scheme}\label{IV}
\begin{figure}[htbp]
    \centering
    \includegraphics[width=0.48\textwidth]{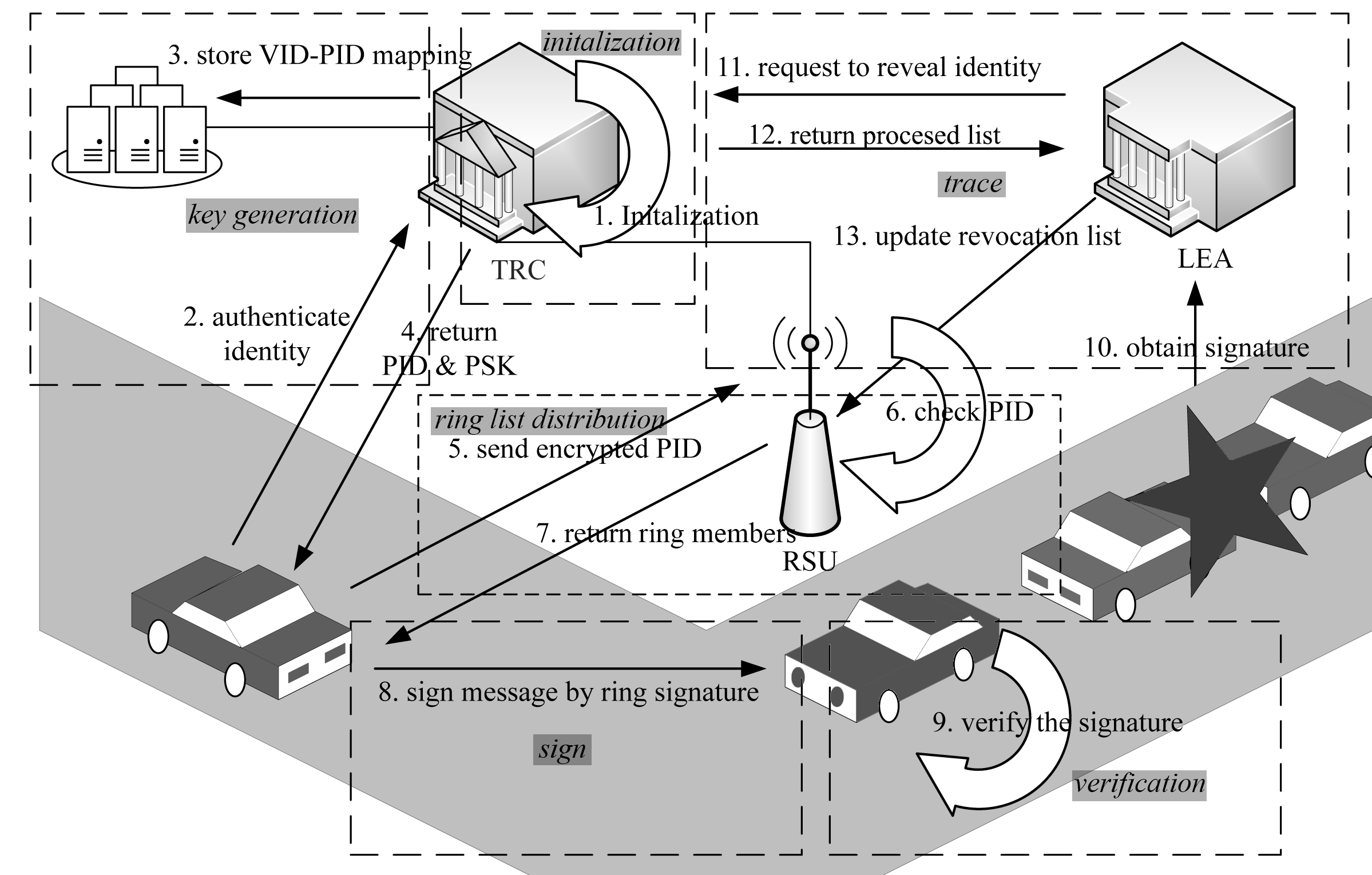}
    \caption{Our proposed scheme}
    \label{fig:03}
\end{figure}

In this section, we illustrate our scheme in detail.
First, we use an abstract pseudonym life-cycle~\cite{petit2014pseudonym} as shown in Fig.~\ref{fig:03} to roughly describe how our scheme works.
We divide the whole life-cycle into six phases: {\em initialization}, {\em key generation}, {\em ring list distribution}, {\em sign}, {\em verification} and {\em trace}.

After the initialization of TRC, each vehicle or RSU is assumed to obtain a pair of the public key and private key from TRC. 
The public key in this system is also regarded as the corresponding pseudonym.
RSU also obtains a set of vehicles' pseudonyms from TRC in the progress of the key generation.

Once a vehicle enters into a certain region,  it obtains a fresh pseudonym list from the local RSU by sending its pseudonym.
Then the vehicle can choose pseudonyms from the list to sign messages by identity-based ring signature.
Other vehicles can verify the signature locally by using the public parameters pre-loaded in OBUs.
Moreover, LEA can reveal the real identity with the help of TRC in some specific scenarios (typically for misbehaviors).
RSUs update the pseudonym revocation list (PRL) from LEAs for filtering the requests from invalid pseudonyms.
Relevant notations are listed in Table~\ref{tab:01}.

\subsection{Initialization}
In the beginning, for the given security parameter $\kappa$, TRC chooses the master secret key $s\in_R\mathbb{Z}_q^*$ randomly and outputs $\{\mathbb{G}_1,\mathbb{G}_2,\mathbb{G}_T,P,Q,PK_1,PK_2,q,\hat{e},H_1,H_2,H\}$ as illustrated in Section~\ref{III-B}.
Then LEA chooses its private key $s_{trac}\in_R\mathbb{Z}_q^*$ randomly, and calculates the corresponding public key $PK_{trac}=s_{trac}\cdot Q$.
Finally, TRC sets the public parameters as $\mathsf{PP}=\{\mathbb{G}_1,\mathbb{G}_2,\mathbb{G}_T,P,Q,PK_1,PK_2,PK_{trac},q,\hat{e},H_1,H_2,H\}$.
We emphasize that the pairing used in the proposed scheme is asymmetric, i.e., $\mathbb{G}_1\neq\mathbb{G}_2$ (this kind of pairing is sometimes called type 3, e.g., see~\cite{galbraith2008pairings}).

\begin{table}[htbp]
\caption{Notation declarations}
\begin{center}
\begin{tabular}{|c|c|}
\hline
\textbf{Notations}& \textbf{Explanation}\\
\hline
$s$ & The master secret key \\
\hline
$s_{trac}$ & The private key of LEA \\
\hline
$PK_{trac}$ & The public key of LEA \\
\hline
$\mathsf{PP}$ & Public parameters \\
\hline
$\mathsf{VID}_i$& Real ID number of vehicle $i$ \\
\hline
$\mathsf{PID}_i$& Public key (pseudonym) of vehicle $i$ \\
\hline
$\mathsf{PSK}_i$& Private key of vehicle $i$ \\
\hline
$\mathsf{RID}_j$& Public key of RSU $j$\\
\hline
$\mathsf{RSK}_j$& Private key of RSU $j$\\
\hline
$K_{i-j}$ & A shared secret key between vehicle $i$ and RSU $j$ \\
\hline
$L$ & The ring list generated by RSUs \\
\hline
$t_d$ & The expired date of $L$ \\
\hline
$L_s$ & The ring list used in ring signature\\
\hline
$t$ & The timestamp for signature \\
\hline
$tag$ & The traceable tag for signature \\
\hline
$\mathsf{KeyGen}(\cdot)$ & The key generation algorithm in Section~\ref{III-C} \\
\hline
$\mathsf{Sign}_{sk}(\cdot)$ & The ring signature algorithm in Section~\ref{III-C}\\
\hline
$\mathsf{Verify}_{L_s}(\cdot)$ & The verification algorithm in Section~\ref{III-C}\\
\hline
$\mathsf{Enc}_{pk}(\cdot)$ & The public encryption algorithm in Section~\ref{III-C}\\
\hline
$\mathsf{Dec}_{sk}(\cdot)$ & The public decryption algorithm in Section~\ref{III-C}\\
\hline
$\mathcal{ENC}_{k}(\cdot)$ & A symmetric encryption algorithm (e.g., AES) \\
\hline
$\mathcal{DEC}_{k}(\cdot)$ & A symmetric decryption algorithm corresponding to $\mathcal{ENC}$\\
\hline
$\mathcal{HMAC}_{k}(\cdot)$ & A symmetric hash-based message authentication code \\
\hline
$a||b$ & String concatenation of a and b \\
\hline
$\mathsf{len}(\cdot)$& Return the number of items in an object \\
\hline
$PRL$ & Pseudonym revocation list\\
\hline
\end{tabular}
\label{tab:01}
\end{center}
\end{table}

\subsection{Key generation}

After the initialization, vehicles, and RSUs obtain key pairs from TRC and pre-load $PP$.
For a vehicle with its real identity $\mathsf{VID}_i$, TRC runs $\mathsf{KeyGen}(\mathsf{VID}_i)$ and sends $(\mathsf{PID}_i,\mathsf{PSK}_i,\mathsf{PP})$ to this vehicle.
These parameters $(\mathsf{PID}_i,\mathsf{PSK}_i,\mathsf{PP})$ are pre-loaded into the tamper-proof device in OBUs.
For an RSU $j$ with its real identity $\mathsf{ID}_j$, similarly, it obtains the key pair $(\mathsf{RID}_j,\mathsf{RSK}_j)$ from TRC, i.e. $\mathsf{RSK}_j = s\cdot \mathsf{RID}_j$.
The complete procedure is shown in Table~\ref{tab:gen}.

\begin{table}[htbp]
\caption{The procedure of key generation}
\center
\label{tab:gen}
\begin{tabular}{|l|}
\hline
\textbf{Key generation:}\\
\begin{minipage}[t]{\linewidth-1cm}
If $id$ belongs to OBU, then
  \begin{itemize}
    \item[1.] $\mathsf{VID}:=id$
    \item[2.] $\mathsf{PID}=H_1(\mathsf{VID})$
    \item[3.] $\mathsf{PSK}=s\cdot \mathsf{PID}$
    \item[4.] record $\{\mathsf{VID}:\mathsf{PID}\}$ mapping
  \end{itemize}
If $id$ belongs to RSU, then
  \begin{itemize}
    \item[1.] $\mathsf{RID}=H_2(id)$
    \item[2.] $\mathsf{RSK}=s\cdot \mathsf{RID}$
  \end{itemize}
\vspace{0.1cm}
\end{minipage}\\
\hline
\end{tabular}
\end{table}

\subsection{Ring list distribution}

To illustrate this procedure, assume that a vehicle $i$ is communicating with an RSU $j$.
In this process, RSU $j$ always broadcasts $\mathsf{RID}_j$ in a designated area.
When vehicle $i$ receives $\mathsf{RID}_j$, it will request the ring list $L$ from RSU $j$ by sending its encrypted public key $\mathsf{PID}_i$.

Here the identity-based encryption algorithm $\mathsf{Enc}_{pk}(\cdot)$ is adopted to encrypt $\mathsf{PID}_i$.
The findings in~\cite{galbraith2008pairings} show that there is an efficient way to transform the elements of $\mathbb{G}_1$ to a short representation in such an asymmetric pairing setting.
Therefore, let $\mathsf{PID}_i'$ be the short representation of $\mathsf{PID}_i$, then $\mathsf{Enc}_{\mathsf{RID}_j}(\mathsf{PID}_i)$ and $\mathsf{Dec}_{\mathsf{RSK}_j}(C)$ are described in Table~\ref{tab:algo}.

\begin{table}[htbp]
\caption{The algorithms of $\mathsf{Enc}$ and $\mathsf{Dec}$}
\center
\label{tab:algo}
\begin{tabular}{|l|}
\hline
\textbf{Encryption:}\\
\hline
\begin{minipage}[t]{\linewidth-1cm}
\begin{itemize}
  \item[1.] Choose $r\in_R\mathbb{Z}_q^*$ randomly
  \item[2.] Compute $g=\hat{e}(PK_1,\mathsf{RID}_j)$
  \item[3.] Transform $\mathsf{PID}_i$ into $\mathsf{PID}_i'$
  \item[4.] Set ciphertext $C:=(rP,\mathsf{PID}_i'\xor H(g^r))$
\end{itemize}
\vspace{0.1cm}
\end{minipage}\\
\hline
\textbf{Decryption:}\\
\hline
\begin{minipage}[t]{\linewidth-1cm}
\begin{itemize}
  \item[1.] For ciphertext $C=(U,V)$
  \item[2.] Decrypt $C$ by using $\mathsf{RSK}_j$: $\mathsf{PID}_i'=V\xor H(\hat{e}(U,\mathsf{RSK}_j))$
  \item[3.] Restore $\mathsf{PID}_i$ from $\mathsf{PID}_i'$
\end{itemize}
\vspace{0.1cm}
\end{minipage}\\
\hline
\end{tabular}
\end{table}

After that, RSU will check whether $\mathsf{PID}_i$ is in PRL or not.
If not, RSU will return a ring list $L$ for the valid vehicle; otherwise, RSU will reject the request.
More precisely, the procedure is presented in Table~\ref{tab:list}.

\begin{table}[htbp]
\caption{The procedure of ring list distribution}
\center
\label{tab:list}
\begin{tabular}{|l|}
\hline
\textbf{Ring list distribution:}\\
\hline
\begin{minipage}[t]{\linewidth-1cm}
\begin{itemize}
  \item[1.] RSU $j$ broadcasts its public key $\mathsf{RID}_j$.
  \item[2.] When vehicle $i$ receives $\mathsf{RID}_j$, it sends the encrypted pseudonym $C=\mathsf{Enc}_{\mathsf{RID}_j}(\mathsf{PID}_i)$.
  \item[3.] After RSU $j$ receives the ciphertext $C$, it computes $\mathsf{RID}_i = \mathsf{Dec}_{\mathsf{RSK}_j}(C)$.
  \item[4.] RSU checks if $\mathsf{PID}_i$ is in the revocation list $PRL$. If yes, simply rejects the requirement.
  \item[5.] If $\mathsf{PID}_i$ is not in $PRL$, then RSU $j$ computes a shared secret key $K_{j-i}=\hat{e}(\mathsf{PID}_i,\mathsf{RSK}_j)$, and uses Encrypt-then-MAC approach to deliver the ring list $L$, i.e., compute $C^*=\mathcal{ENC}_{K_{j-i}}(L)$ and $\Sigma = \mathcal{HMAC}_{K_{j-i}}(C^*||t_d)$. Finally sends $(C^*,\Sigma,t_d)$ to the vehicle.
  \item[6.] After the vehicle receives the ciphertext, it first computes the shared secret key $K_{i-j}=\hat{e}(\mathsf{PSK}_i,\mathsf{RID}_j)$, then checks the message authentication code $\Sigma$ and recovers $L=\mathcal{DEC}_{K_{i-j}}(C^*)$.
\end{itemize}
\vspace{0.1cm}
\end{minipage}\\
\hline
\end{tabular}
\end{table}

Because of the capability bottleneck of RSUs, we propose that RSUs store the shared keys locally to reduce computation.

\subsection{Sign}
For a vehicle $V_k$ holding a ring list $L$ with an unexpired $t_d$, it first choose $n'-1$ pseudonyms from $L$ randomly to establish a ring list $L_s$, i.e., $L_s=\{\mathsf{PID}_1,\mathsf{PID}_2,\ldots,\mathsf{PID}_k, \ldots,\mathsf{PID}_{n'}\}$.
Then it could use ring signature as described in Section~\ref{III-C}, i.e., $\sigma=\mathsf{Sign}_{\mathsf{PSK}_k}(m||tag||t,L_s)$, where $tag=\hat{e}(H_1(VID_\ell||t), PK_{trac})$.
Finally it broadcasts $(m,\sigma,L_s,t,tag)$.
The detailed procedure is illustrated in Table~\ref{tab:02}.

\begin{table}[htbp]
\caption{The procedure of signing}
\center
\label{tab:02}
\begin{tabular}{|l|}
\hline
\textbf{Sign:}\\
\hline
\begin{minipage}[t]{\linewidth-1cm}
If $t_d$ is not expired, then
  \begin{itemize}
    \item[1.] Choose $n'-1$ $PID$ from $L$ randomly, and set $L_s:=\{PID_1,,PID_2,\ldots,PID_k,\ldots,PID_{n'}\}$
    \item[2.] For $i$ from $1$ to $n'$ and $i\neq k$, compute:\\
    $U_i\leftarrow_R\mathbb{G}_1$, $h_i=H(m||tag||t||L_s||U_i)$
    \item[3.] $r'\leftarrow_R\mathbb{Z}_q^*$, $U_k=r'PID_k-\sum_{r=1,r\neq k}^{n'}(U_i+h_iPID_i)$
    \item[4.] $h_k=H(m||tag||t||L_s||U_k)$, $V=(h_k+r')PSK_k$
    \item[5.] Return $\sigma:=(\{U_i\}_{i=1}^{n'},V)$
  \end{itemize}
\vspace{0.1cm}
\end{minipage}\\
\hline
\end{tabular}
\end{table}

\subsection{Verification}

When the vehicle, e.g., $V_\ell$ receives $(m,\sigma,L_s,t,tag)$, it first checks $t$ to prevent replay attack.
If the check passes, then it runs $\mathsf{Verify}_{L_s}(m||tag||t,\sigma)$ to check the validity of the signature.

As mentioned above, consider the limited capacity of OBUs, we suggest adopting batch verification proposed in~\cite{ferrara2009practical}.
Table.~\ref{tab:03} presents the procedures of single verification and batch verification in detail.

\begin{table}[htbp]
\caption{The procedures of verification}
\center
\label{tab:03}
\begin{tabular}{|l|}
\hline
\textbf{Single verification:}\\
\hline
\begin{minipage}[t]{(\linewidth-1cm)}
\begin{itemize}
  \item[1.] Let $\sigma=(\{U_i\}_{i=1}^{n'},V)$, $L_s=(\mathsf{PID}_1,\mathsf{PID}_2,\ldots,\mathsf{PID}_{n'})$
  \item[2.] For $i$ from $1$ to $n'$, compute:\\
    $h_i=H(m||tag||t||L_s||U_i)$
  \item[3.] Return $\hat{e}(\sum_{i=1}^{n'}(U_i+h_i\mathsf{PID}_i),PK_2)\overset{?}{=}\hat{e}(V,Q)$
\end{itemize}
\vspace{0.1cm}
\end{minipage}\\

\hline
\textbf{Batch verification:}\\
\hline
\begin{minipage}[t]{(\linewidth-1cm)}
\begin{itemize}
  \item[1.] Given $\eta$ messages and corresponding signatures, i.e., $\sigma_{batch}=\{\sigma_1, \sigma_2,\ldots,\sigma_\eta\}$, $M_{batch}=\{m_1,m_2,\ldots,m_{\eta}\}$, and $L_{batch}=\{L_{s1},L_{s2},\ldots,L_{s\eta}\}$
  \item[2.] 
  For $i$ from 1 to $\eta$, and $\forall j\in \{1, \mathsf{len}(L_{si})\}$ , compute:\\
            $h_{ij}=H(m_i||tag_i||t_i||L_{si}||U_{ij})$
  \item[3.] Return \\$\hat{e}(\sum_{i=1}^{\eta}\sum_{j=1}(U_{ij}+h_{ij}\mathsf{PID}_{ij}), PK_2)\overset{?}{=}\hat{e}(\sum_{i=1}^{\eta}V_i,Q)$
\end{itemize}
\vspace{0.1cm}
\end{minipage}\\
\hline
\end{tabular}
\end{table}

\subsection{Trace}

Once LEA detects misbehaviors in VANETs, it first calculates $tag'=tag^{1/s_{trac}}$ and sends $\{L_s$,$t\}$ to TRC.
For $L_s=\{\mathsf{PID}_1,\mathsf{PID}_2,\ldots,\mathsf{PID}_{n'}\}$, TRC computes $H_i'=\hat{e}(H_i(\mathsf{VID}_i||t),P)$, $\forall i\in\{1,2,\ldots,n'\}$, and returns $\cup_{i=1}^{n'}\{H_i'\}$.
By comparing $tag'$ and $\cup_{i=1}^{n'}\{H_i'\}$, LEA can determine the signer's pseudonym.
Furthermore, LEA would find out the signer's true identity by sending the pseudonym to TRC. 

\section{Security analysis}\label{V}
\begin{figure}[htbp]
    \centering
    \includegraphics[width=\linewidth]{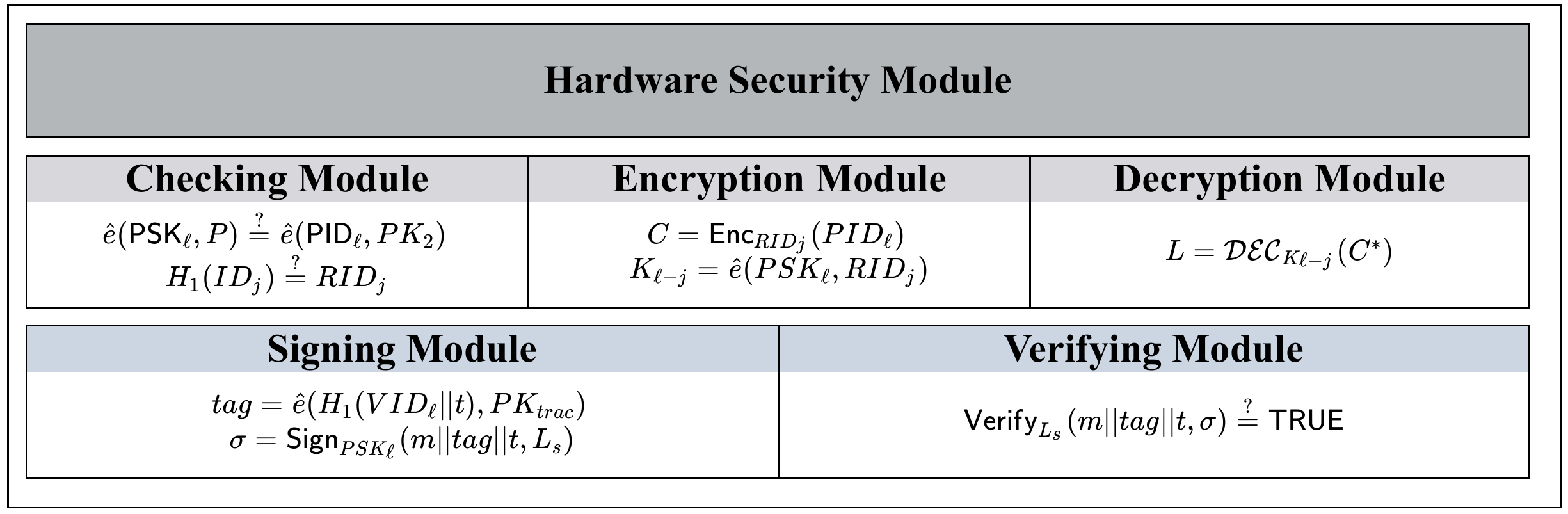}
    \caption{Abstract construction of HSM}
    \label{fig:04}
\end{figure}

In this section, we analyze the security of our scheme.
Recalling the assumptions in Section~\ref{III}, each OBU has a secure module called HSM, so that HSM provides an independent environment to perform these required cryptographic operations.
Each HSM consists of 5 sub-modules as shown in Fig~\ref{fig:04}.
Hereafter, all analysis is based on the assumption of HSMs.

\subsection{Correctness}

In V2I communication, when vehicle $\mathsf{V}_\ell$ enters the region within the range of $\mathsf{RSU}_j$, it will receive the broadcasting $\mathsf{RID}_j$ in this region.
Once $\mathsf{V}_\ell$ obtains $\mathsf{RID}_j$, it can invoke the Checking Module to check the validity of $\mathsf{RSU}_j$.
In the process of delivering $\mathsf{PID}_\ell$, the correctness and security are guaranteed by the property of identity-based encryption scheme~\cite{boneh2001identity}.
According to the property of bilinear pairing, we know that:
\begin{equation*}
\begin{split}
K_{\ell-j}&=\hat{e}(\mathsf{PSK}_\ell,\mathsf{RID}_j) \\
&=\hat{e}(\mathsf{PID}_\ell,\mathsf{RID}_j)^s  \\
&=\hat{e}(\mathsf{PID}_\ell,\mathsf{RSK}_j) \\
&=K_{j-\ell}\\
\end{split}
\end{equation*}
It is clearly that both $\mathsf{RSU}_j$ and $\mathsf{V}_\ell$ obtain the same shared key, and can establish an efficient trusted channel for further communication through symmetric cryptography.

In V2V communication, if the procedure of signing a message $m$ is executed correctly, then the corresponding signature $\sigma$ must satisfy the verifying equation.

\subsection{Unforgeability}

We say that if a signature $\sigma$ is unforgeable, an adversary should not be able to generate a signature for a new message, given a few signatures corresponding to the messages of his own choice.
Since the ring signature scheme~\cite{chow2005efficient} we employed is proven to be unforgeable against chosen message attacks in the random oracle model, we note that our protocol is unforgeable.

\subsection{Conditional anonymity}

In V2I communication, RSU $j$ only knows the pseudonym of $\mathsf{V}_\ell$ rather than the true identity of $\mathsf{V}_\ell$ (i.e.,$\mathsf{VID}_\ell$).
As for V2V communication, the true signer is hidden in a set of pseudonyms $L$.
For any eavesdroppers in VANETs, they cannot figure out the true signer from $L$ even though they knowing the corresponding $tag$.
Only the LEA can identify the signer in $L$ through the secret tracing key $s_{trac}$.

\subsection{Against reply attack}

Since each message contains a timestamp in our scheme, namely, once the vehicles figure out that the message is expired, then it will be abandoned before being verified.
If an adversary forges a fresh timestamp to replace the original one, then this message must not be able to pass the verification.

\section{Performance analysis}\label{VI}
In this section, we evaluate the performance of the proposed scheme in terms of both computation cost and communication cost.
To be specific, we adopt ``MNT159'' with degree 6 as the asymmetric group $\mathbb{G}_1$ which has a 159-bit base field size.
In contrast to some recent work where symmetric groups like ``SS512'' are employed, we argue that ``MNT159'' has a shorter presentation for group elements and is more efficient in batch verification according to~\cite{ferrara2009practical} while providing an approximate secure level~\cite{dingledine2009financial}.
Table~\ref{tab:0y} lists the concrete security level under different elliptic-curve settings.

\begin{table}[htbp]
    \centering
    \begin{threeparttable}
    \caption{The security level of different elliptic curves}
    \begin{tabular}{|c|c|c|c|c|}
        \hline
        & $\mathbb{G}_1$ & $\mathbb{G}_2$ & $\mathbb{G}_T$ & Security level \\
        \hline
        MNT159 & 159 bits & 477 bits & 945 bits & 70 bits \\
        \hline
        SS512 & 512 bits & 512 bits & 1024 bits & 80 bits \\
        \hline
        Secp160k1\tnote{*} & 160 bits & - & - & 80 bits\\
        \hline
    \end{tabular}
    \begin{tablenotes}
        \footnotesize
        \item[*] Secp160k1 is a standardized elliptic curve which is widely used in mainstream applications such as SSL. 
    \end{tablenotes}
    \label{tab:0y}
    \end{threeparttable}
\end{table}

All experiments are performed on Raspberry Pi 3b+ which is a cheap microcomputer with a 1.4 GHz ARM CPU and 1 GB RAM running Debian Linux operation system.
Besides, we invoke the cryptographic framework CHARM~\cite{charm13} to
implement the proposed scheme.

\subsection{Computation cost}

To illustrate the computation cost of our scheme clearly, we provide the real-world benchmark in CHARM of different operations in three types of elliptic curves as shown in Table~\ref{tab:0x}.

\begin{table*}[htbp]
    \centering
    \begin{threeparttable}
    \caption{The time cost\tnote{*} of executing different cryptographic operations in different Elliptic Curves}
    \begin{tabular}{|c|c|c|c|c|}
        \hline
        Symbol & Description & MNT159 & SS512 & Secp160k1 \\
        \hline
        $T_{bp}$ & The execution time of one bilinear pairing operation & 117.10 & 54.50 & -\\
        \hline
        $T_{ep}$ & The execution time of one exponentiation operation on $\mathbb{G}_T$ & 10.82 & 2.06 & -\\
        \hline
        $T_{em}$ & The execution time of one scale multiplication operation on $\mathbb{G}_1$ & 3.49 & 9.94 & 3.32 \\
        \hline
        $T_{mph}$ & The execution time of one map-to-point hash operation on $\mathbb{G}_1$ & 0.446 & 31.91 & -\\
        \hline
    \end{tabular}
    \begin{tablenotes}
        \footnotesize
        \item[*] The execution time is measured in milliseconds, each operation is evaluated in 1000 times and calculate the average value.
    \end{tablenotes}
    \label{tab:0x}
    \end{threeparttable}
\end{table*}

\begin{table*}[htbp]
    \centering
    \begin{threeparttable}
    \caption{The computation cost of signing and verification for a single message\tnote{*}}
    \begin{tabular}{|c|c|c|}
    \hline
        & Message signing (ms) & Message verification (ms) \\
    \hline
    Jiang et al.'s scheme~\cite{jiang2014anonymous}  & $5T_{em}\approx16.6$ & $6T_{em}\approx19.92$ \\
    \hline
    Zeng et al.'s scheme~\cite{zeng2015privacy} & $3T_{bp}+4T_{ep}+4T_{em}+2T_{mph}\approx275.02$ & $3T_{bp}+3T_{ep}+4T_{em}\approx209.14$ \\
    \hline
    Our scheme & $3T_{em}+T_{bp}\approx127.57$ & $2(T_{bp}+T_{em})\approx255.84$ \\
    \hline    
    \end{tabular}
    \begin{tablenotes}
        \footnotesize
        \item[*] The ring size is 2.
    \end{tablenotes}
    \end{threeparttable}
    \label{tab:08}
\end{table*}

\begin{figure}[htbp]
   \centering
   \includegraphics[width=\linewidth]{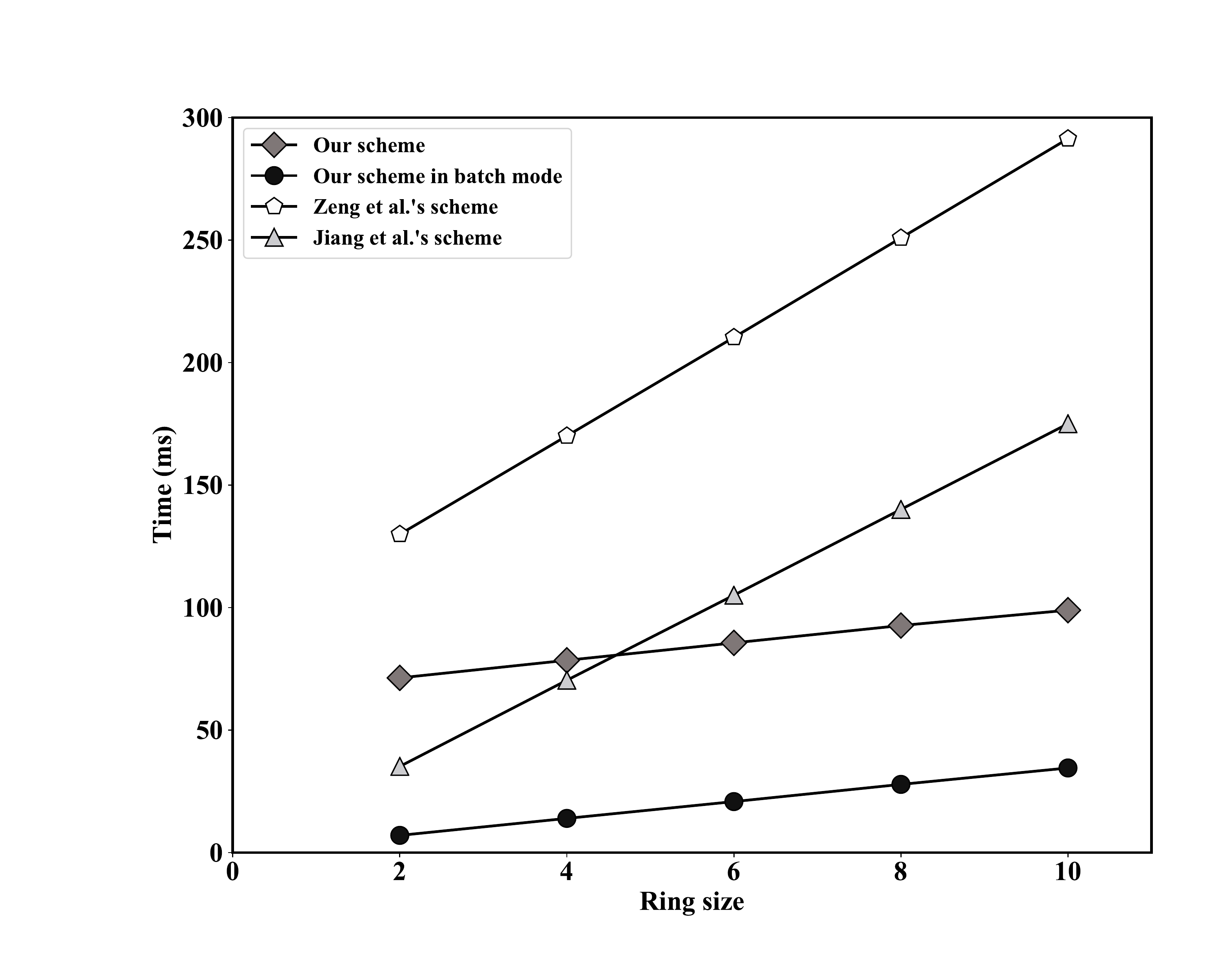}   
   \caption{The computation cost of verification with respect to ring size}
   \label{fig:05}
\end{figure}

\begin{figure}[htbp]
   \centering
   \includegraphics[scale=0.5]{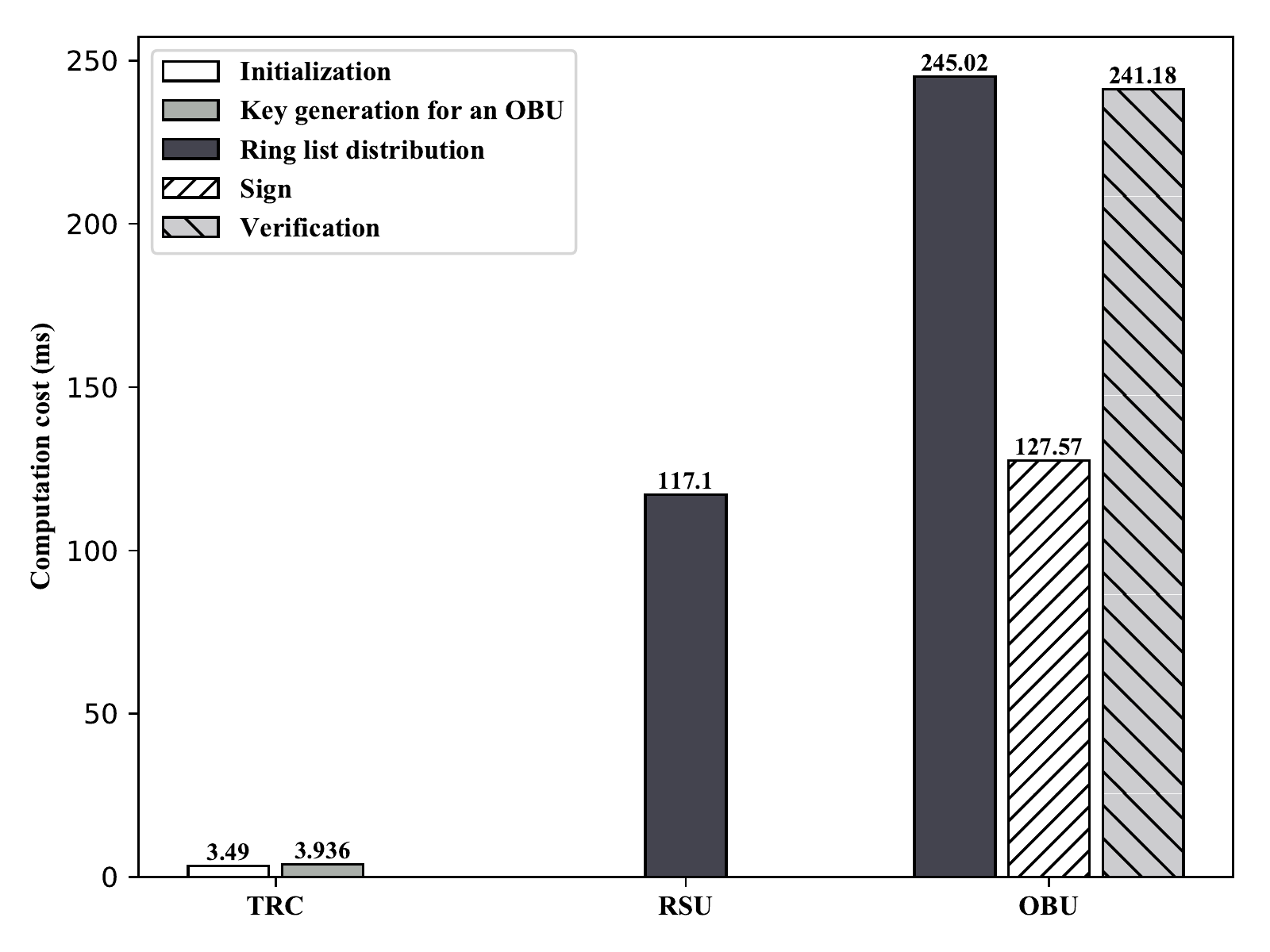}
   \caption{The computation cost for each phase and entity}
   \label{fig:06}
\end{figure}

\begin{figure*}[htbp]
   \centering
   \includegraphics[width=18cm]{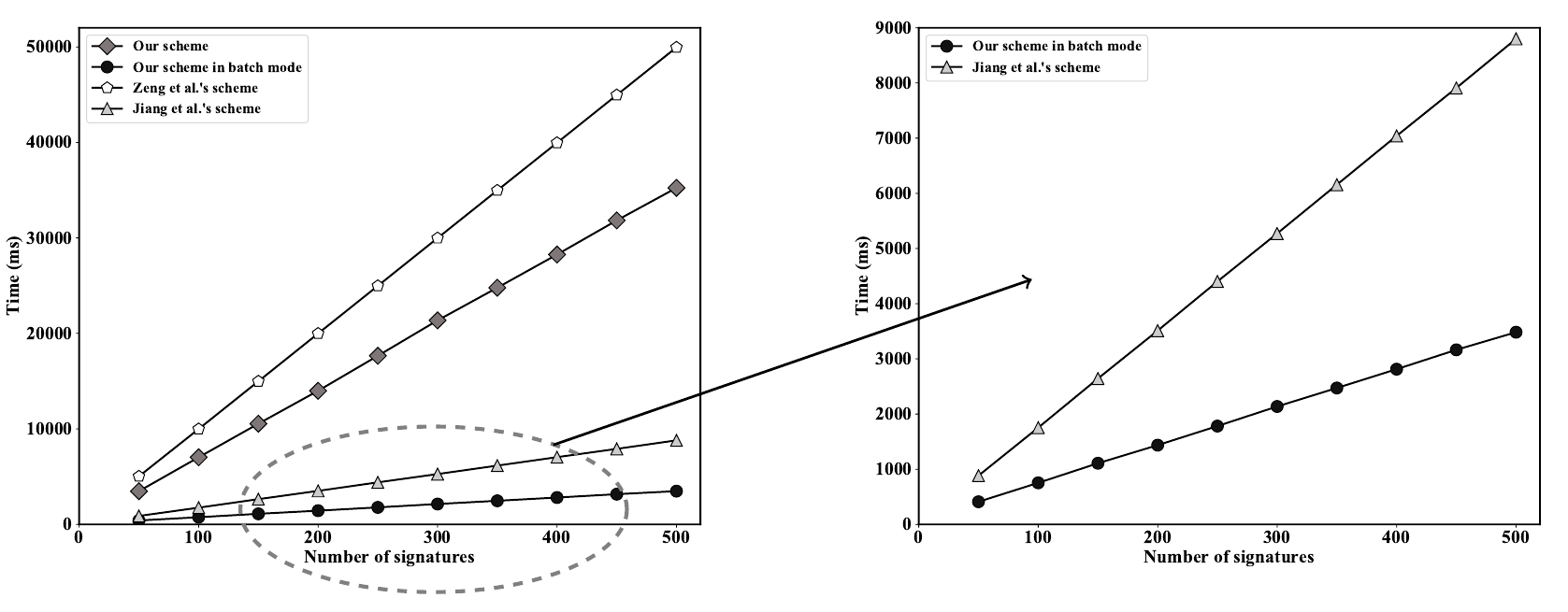}
   \caption{The computation cost of verification with respect to the number of messages (ring size = 2)}
   \label{fig:07}
\end{figure*}


Furthermore, to present the performance differences, we compare the computation cost of verification of our scheme, and the other two schemes proposed by Zeng et al.~\cite{zeng2015privacy}, Jiang et al.~\cite{jiang2014anonymous} respectively.
Zeng et al.'s scheme is based on symmetric bilinear pairing, while Jiang et al.'s scheme is based on a normal elliptic curve.
So that the measurements for these two schemes are taken on SS512 and Secp160k1 respectively.

Though it seems that our scheme should an employ elliptic curve like Secp160k1 rather than MNT159 according to Table~\ref{tab:0x}, the bilinear pairings have advantages in building identity-based cryptosystem which can reduce the communication overhead in high-density networks.

It is widely assumed~\cite{ali2019authentication,safi2018cloud,petit2014pseudonym} that OBUs have limited computation capacity.
However, Fig.~\ref{fig:05} shows the computation cost of the ring signature depends on the number of ring members.
Since a 2-member ring is applied in some lightweight applications such as email signing~\cite{adida2006lightweight} and VANETs~\cite{zeng2018concurrently}, we argue that privacy can be assured with using the 2-member ring in our scheme.
Based on the benchmark, when ring size is 2, the computation cost for each phase and entity is shown in Fig.~\ref{fig:06}.

According to the implementation results, we find that distributing a ring list is time-consuming.
However, in the phase of ring-list distribution, the shared key could be cached in OBU which will avoid double-counting in the next time.

The results in Table~\ref{tab:08} imply that the schemes based on bilinear pairings have no advantages on the computation of signing and verification procedure.
However, in most scenarios, vehicles receive a batch of messages simultaneously rather than a single message.
The results in Fig.~\ref{fig:07} shows that enabling batch verification will greatly decrease the computation overhead.

\subsection{Communication cost}

\begin{table}[htbp]
    \centering
    \resizebox{\linewidth}{!}{
    \begin{threeparttable}
    \caption{Communication cost (Bytes)\tnote{*}}
    \begin{tabular}{|c|c|c|}
    \hline
       & A single pseudonym & A single signature \\
    \hline
    Jiang et al.'s scheme~\cite{jiang2014anonymous}  & 38 & 308 \\
    \hline
    Zeng et al.'s scheme~\cite{zeng2015privacy} & 90 & 936\\
    \hline
    Our scheme & 30 & 90\\
    \hline
    \end{tabular}
    
    \begin{tablenotes}
        \footnotesize
        \item[*] The communication cost is evaluated by the built-in function $\mathsf{serialize()}$ with enabling compression in CHARM.
    \end{tablenotes}
    \label{tab:09}
    \end{threeparttable}
    }
\end{table}

Because the approaches in Jiang et al.'s scheme~\cite{jiang2014anonymous} is based on the ElGamal cryptosystem where the generated signature consists of group elements in pairs.
On the other hand, the generated signature based on bilinear pairing could consist of one group element (e.g., short signature).

Zeng et al.'s scheme is based on symmetric bilinear pairing where the size of elements in $\mathbb{G}_T$ is much larger than that in $\mathbb{G}_T$ under the approximate security setting.
The results in Table~\ref{tab:09} also show that our scheme is more efficient in terms of communication costs than the other ring signature-based schemes.

\section{Conclusion}\label{VII}
In this paper, we propose an efficient identity-based batch verification scheme for VANETs based on the ring signature.
Unlike other ring signature-based schemes, we restrict the generation of a ring to avoid disruptions from malicious vehicles.
Consider that VANETs are highly dense in most real-world scenarios, we adopt batch verification and bilinear pairing to reduce the computation and communication cost respectively.
To simulate the environment of OBUs, we implement the proposed scheme on the Raspberry Pi 3b+ platform.
By comparing with other related schemes, our scheme is much more efficient in computation and communication with batch mode.

Due to limited space, the approach for finding invalid signatures in batch verification is not shown in this paper.
As a possible direction of future work, we will give a more comprehensive illustration of applying batch verification in VANETs.
Moreover, it might be interesting to consider building HSMs in real-world applications based on the trusted execution environment (TEE) such as ARM's TrustZone.

\vspace{12pt}
\end{document}